\documentclass[prl,twocolumn]{revtex4-1}
\usepackage{graphicx}
\usepackage{color}
\begin{document}
\title{Validity of compressibility equation and
  Kirkwood-Buff theory in crystalline matter}
\author{Peter Kr\"uger}
\email{pkruger@chiba-u.jp}
\affiliation{Graduate School of Engineering
  and Molecular Chirality Research Center,
  Chiba University, Chiba 263-8522, Japan}

\begin{abstract}
Volume integrals over the radial pair-distribution function,
so-called Kirkwood-Buff integrals (KBI)
play a central role in the theory of solutions,
by linking structural with thermodynamic information.
The simplest example is the compressibility equation, a fundamental
relation in statistical mechanics of fluids.
Until now, KBI theory could not be applied to crystals,
because the integrals strongly
diverge when computed in the standard way.
We solve the divergence problem and generalize KBI theory to crystalline
matter by using the recently proposed finite-volume theory.
For crystals with harmonic interaction,
we derive an analytic expression for the peak shape of the pair-distribution function
at finite temperature. From this we demonstrate that the compressibility equation holds exactly in
harmonic crystals.
\end{abstract}




\maketitle
In statistical mechanics a close link exists between fluctuations of extensive variables
and response functions. For example, the density and concentration fluctuations in a fluid
are directly related to the isothermal compressibility, partial molar volumes and
diffusion coefficients~\cite{mcquarrie,bennaim,guevara20}.
Density fluctuations can be accessed experimentally
as the long-wave-length limit of the structure factor and 
theoretically from volume integrals over the pair-distribution function (PDF),
so-called Kirkwood-Buff integrals (KBIs)~\cite{kb52}.
KBI theory was developed for fluids, i.e. homogeneous and isotropic 
systems. In solids, these symmetries are sponspontaneously broken
and statistical mechanics commonly starts from the perfect crystal~\cite{landau}.
Defects and lattice vibrations are treated as a perturbation~\cite{ashcroft}.
This approach is efficient at low temperature and low defect
concentration, but it breaks down near the solid-liquid phase transition
and in amorphous and nanostructured materials.
Therefore a common set of statistical mechanical techniques, valid for
all states of matter is highly desirable.
The PDF should be a key element of such a framework, since it can be easily obtained
for all kinds of matter~\cite{billinge07}, including amorphous
and crystalline solids~\cite{toby,proffen05}, liquids and gases.
This raises the question whether KBI theory,
which is a cornerstone of the theory of solutions~\cite{bennaim,kb52,shulgin06,milzetti18},
is valid in the solid state.
Surprisingly, this basic question has, to the best of our knowledge,
not yet been answered.

In this letter, we demonstrate that KBI theory is valid for crystalline
matter, i.e. thermodynamic quantities such as the isothermal compressibility,
can be obtained from the PDF, in principle in the same way as in fluids.
In crystals, KBIs diverge when computed in the standard way as ``running''
integrals. We solve the divergence problem by starting from
the recently developed KBI theory for finite volumes~\cite{kruger13}.
For a monoatomic crystal with harmonic interactions,
we derive an analytic expression for the thermal broadening of the PDF.
From this we calculate the crystal KBI at finite temperatures
and demonstrate that the compressibility equation holds exactly in
harmonic crystals. Our findings extend the validity and applicability of KBI theory,
hitherto used only in homogeneous fluids, to all phases of matter,
and thus open up new ways for the thermodynamic modeling of complex materials.

We consider a monoatomic system for simplicity. Generalization of the
present theory to multicomponent systems is straightforward.
The instantaneous position of particle~$i$ is ${\bf r}_i$. 
The single-particle density is given by
$\rho({\bf r})=\langle\sum_{i}\delta({\bf r}-{\bf r}_i)\rangle$,
where $\langle \dots \rangle$ denotes the grand-canonical ensemble average.
The PDF is defined as
\begin{equation}\label{grr'}
  g({\bf r}',{\bf r}'')
  =\frac{ \langle\sum_{i\ne j}\delta({\bf r}'-{\bf r}_i)
   \delta({\bf r}''-{\bf r}_j)\rangle}{\rho({\bf r}')\rho({\bf r}'')} \;.
\end{equation}
For a homogeneous and isotropic system
the density is constant $\rho({\bf r})=\rho$. The PDF depends only on
the pair distance
$r=|{\bf r}|=|{\bf r}'-{\bf r}''|$ and simplifies to
\begin{equation}\label{gr}
  g(r)
  = \frac{1}{V\rho^{2}}\langle\sum_{i\ne j}
  \delta({\bf r}-{\bf r}_i+{\bf r}_j)\rangle \;,
\end{equation}
where $V$ is the volume of the system.
In a crystal, the translational and rotational symmetry are broken,
and so Eq.~(\ref{gr}) does not follow directly from Eq.~(\ref{grr'}).
Here we consider the statistical ensemble of arbitrarily shifted
and rotated crystals, corresponding to a powder sample~\cite{kruger20}.
Since this ensemble is homogeneous and isotropic, Eq.~(\ref{gr}) is valid
and commonly used in powder diffraction analysis~\cite{toby}.
The finite-volume KBI is defined as~\cite{kruger13}
\begin{equation}\label{GV}
  G^V =  \frac{1}{V} \int_V d{\bf r}'\int_V d{\bf r}'' 
 ( g(r)-1 ) \;.
\end{equation}
By inserting Eq.~(\ref{gr}) into (\ref{GV}) we obtain
the well-known relation between the KBI and the particle number fluctuations
in the volume~$V$~\cite{kb52,kruger13}
\begin{equation}\label{gamma}
  \Gamma^{-1}(V) \equiv
  \frac{\langle N^2\rangle -\langle N\rangle^2 } {\langle N\rangle}
= 1 + \rho G^V \;,
\end{equation}
where $N$ is the instantaneous number of particles in~$V$
and $\langle N\rangle=\rho V$ is its grand-canonical ensemble average.
We shall refer to $\Gamma^{-1}$ as the fluctuation function.
It is also known as the thermodynamic correction factor~\cite{schnell11}.

\begin{figure}
 \begin{center}
\includegraphics[width=1.\columnwidth]{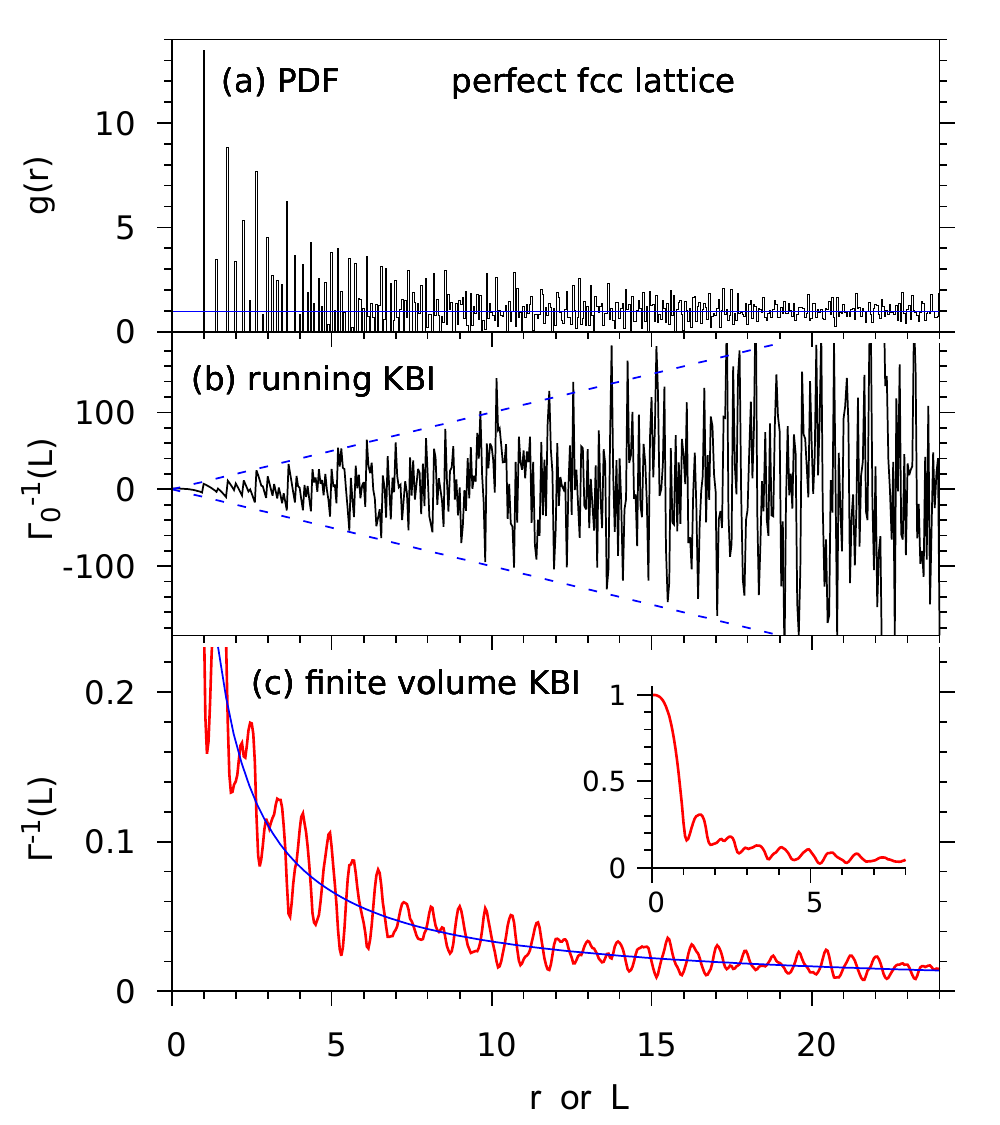}
\caption{
  PDF $g(r)$ and correlation function~$\Gamma^{-1}(L)$,
  Eq.~(\protect\ref{gamma}),
  of a perfect fcc lattice.
  $r$ is the pair distance 
  and $L$ is the upper integration bound.
  The nearest neighbor distance~$d$ is taken as unit length.
  (a) PDF as histogram plot with bin size $\Delta r=0.05$.
  (b) $\Gamma^{-1}_0(L)$ computed with running KBI,
  Eq.~(\protect\ref{G0L}).
  (c) $\Gamma^{-1}(L)$ computed with finite-volume KBI,
  Eq.~(\protect\ref{GL}).
   The inset shows $\Gamma^{-1}(L)$ on the full $y$-scale.
   The blue lines are a guide to the eye.
   They are $y=1$ in (a), $y=\pm 10L$ in (b), $y=0.33/L$ in (c).
}\label{fig1}
\end{center}
\end{figure}

We first consider a perfect crystal at zero temperature
neglecting zero-point motion.
The PDF in Eq.~(\ref{gr}) becomes
\begin{equation}\label{gcrystal0}
4\pi r^2\rho\; g(r) =\sum_{{\bf R}\ne {\bf 0}}\delta(r-|{\bf R}|)
=\sum_{s=1}^\infty n_s \delta(r-R_s) \;,
\end{equation}
where ${\bf R}$ are the lattice sites and
the sum on the r.h.s. runs over shells~$s$,
containing $n_s$ lattice points at a distance $R_s$ from the origin.
Throughout this paper, the nearest neighbor distance~$d$ is taken as
the unit length. The PDF of the fcc lattice is shown in Fig.~\ref{fig1}~a.
It compares well with low temperature PDF data of fcc crystals,
such as Al~\cite{toby}.
In the limit $V\rightarrow\infty$, Eq.~(\ref{GV}) simplifies to~\cite{kb52}
\begin{equation}\label{ginf}
  G^\infty = \int_0^\infty (g(r)-1) 4\pi r^2 dr \;,
\end{equation}
provided that the integrals in Eq.~(\ref{GV}) converge absolutely.
This condition is met in fluids, since $g(r)-1$ vanishes for distances larger
than the correlation length. KBIs are commonly computed as running integrals,
\begin{equation}\label{G0L}
  G_0(L) = \int_0^L (g(r)-1) 4\pi r^2 dr \;,
\end{equation}
where $L$ is a cut-off radius.
The corresponding fluctuation function
$\Gamma^{-1}_0(L) = 1 + \rho G_0(L)$, is plotted in Fig.~\ref{fig1}~b.
It shows large oscillations whose amplitude grows linearly with~$L$.
Comparing Figs~\ref{fig1}~a and \ref{fig1}~b,
we see that the oscillations of the PDF
become strongly amplified upon volume integration. This effect may also
occur in liquids for small~$L$~\cite{kruger13}.
However, in liquids the running KBI eventually converges, namely when
$L$ exceeds the correlation length.
In a crystal, the correlation length is infinite and
the running KBI never converges.
From this result, one might think that KBI theory cannot be
applied to crystals. We now show that this is not true and
we devise a method for calculating KBIs in crystals.

The finite volume KBI in Eq.~(\ref{GV}) can be transformed exactly
to a one-dimensional integral~\cite{kruger13,dawass18} as
\begin{equation}\label{GL}
  G^V = G(L) = \int_0^L (g(r)-1) y(r/L) 4\pi r^2 dr \;,
\end{equation}
where $y(x)$
is a geometrical function characteristic of the shape of the volume~$V$
and $L$ is the maximum distance between any two points in~$V$.
We consider a sphere of diameter~$L$. In this case,
$y(x) = 1- 3x/2 + x^3/2$, $x=r/L$~\cite{kruger13}.

The fluctuation function~$\Gamma^{-1}(L)$ obtained with the
finite-volume KBI of Eq.~(\ref{GL}) is shown in Fig.~\ref{fig1}~c.
It clearly converges for $L\rightarrow\infty$,
in sharp contrast to the running KBI (Fig.~\ref{fig1}~b).
We find $\Gamma^{-1}(L)\rightarrow 0$, which is the correct limit,
since the particles are immobile in a perfect crystal and thus the
particle number fluctuations must vanish for $V\rightarrow\infty$.
For a system with finite correlation length, e.g. a fluid away
from the critical point, we previously proved that 
$\Gamma^{-1}(L)-\Gamma^{-1}(\infty)$ varies as $1/L$~\cite{kruger13}.
As seen in Fig.~\ref{fig1}~c, this also holds for crystals, despite
the fact that the particle positions are correlated over infinite
distances.
Importantly, $\Gamma^{-1}(L)$ is strictly positive,
which is a necessary property of a variance (Eq.~\ref{gamma}).
In contrast, $\Gamma_0^{-1}(L)$ (Fig.~\ref{fig1}~b) is not always positive,
and thus does not describe particle
fluctuations for finite~$L$~\cite{dawass18,kruger18}.

At finite temperature, the peaks of the PDF become broadened due to
vibrations of the atoms around their equilibrium positions.
We now derive an analytic
expression for the broadening function of a monoatomic harmonic crystal.
Let  ${\bf R}+{\bf u}$ be the instantaneous position of the atom
at lattice site ${\bf R}$ and ${\bf u}$ the displacement vector.
A lattice vibration is described by
\begin{equation}\label{equ}
{\bf u}_{\bf k}({\bf R},t) = 
{\bf u}_{{\bf k},0} \cos({\bf k}\cdot {\bf R}-\omega_{\bf k} t) \;,
\end{equation}
where ${\bf k}$ is the wave vector, ${\omega_{\bf k}}$ the frequency and
${\bf u}_{{\bf k},0}$ the amplitude.
For each ${\bf k}$ there are three modes with different
polarization~$\lambda$. We first consider any one of them.
The amplitude ${\bf u}_{{\bf k},0}$ is related to the temperature
through the equipartition theorem as
\begin{equation}\label{equipart}
\frac{1}{2}k_BT =  \frac{1}{2}m \omega_{\bf k}^2
\sum_{\bf R} \langle {u}_{\bf k}({\bf R},t)^2\rangle 
=\frac{1}{4}Nm\omega_{\bf k}^2 {u}_{{\bf k},0}^2 \;,
\end{equation}
where $\langle \dots\rangle$ denotes the time average and
$N$ is the number of atoms.
The relative displacement of the atom at site ${\bf R}$ with respect to
the atom at site ${\bf 0}$ is $\Delta {\bf u}_{{\bf k}}({\bf R},t) =
{\bf u}_{{\bf k}}({\bf R},t) - {\bf u}_{{\bf k}}({\bf 0},t)$.
From Eq.~(\ref{equ}), the time average of the square of the relative
displacement is easily found to be
\begin{equation}\label{DeltaukR}
\langle \Delta {u}_{\bf k}({\bf R})^2\rangle =
{u}_{{\bf k},0}^2 (1-\cos{\bf k}\cdot {\bf R})\;.
\end{equation}
In the harmonic approximation the phonons are
independent and so the mean square displacement is additive.
Therefore the total mean square
$\sigma^2\equiv\langle \Delta {u}({\bf R})^2\rangle$
is obtained by summing (\ref{DeltaukR})
over all wave vectors in the Brillouin zone.
This yields
\begin{equation}
\sigma^2=
\frac{2k_BT}{mN}\frac{V}{8\pi^2}\int_{BZ} d{\bf k}
\frac{1-\cos{\bf k}\cdot {\bf R}}{\omega_{\bf k}^2} \;,
\end{equation}
where $V$ is the volume of the crystal. 
Here, the $k$-space volume element
$d{\bf k}=8\pi^3/V$ and Eq.~(\ref{equipart}) have been used.
We now make the Debye approximation, i.e. we assume $\omega_{\bf k} = ck$,
where $c$ is the speed of sound and we replace the Brillouin zone
by a sphere of radius $k_D = (6\pi^2\rho)^{1/3}$ where $\rho=N/V$ is
the atomic density. After a straightforward integration over
the angular coordinates of ${\bf k}$, we obtain 
\begin{equation}\label{sigma2}
\sigma^2=\frac{k_BT}{\rho mc^2}\frac{1}{\pi^2}\left(k_D -\frac{1}{R}
\int_0^{x_D} \frac{\sin x}{x}dx \right) \;,
\end{equation}
where $x_D\equiv k_DR$.
We write $\rho=(\gamma/d)^3$, where $d$ is the nearest neighbor distance
and $\gamma$ is a dimensionless factor
which depends on the lattice type. $\gamma=1$ for simple cubic and
$\gamma=2^{1/6}\approx 1.12$ for close-packed lattices.
We have $x_D=\gamma(6\pi^2)^{1/3}R/d\approx 4R/d$.
As we are interested in the regime $R\gg d$, we can safely
take the limit $x_D\rightarrow\infty$ in the integral in Eq.~(\ref{sigma2}).
The integral then becomes
$\pi/2$ and we obtain
\begin{equation}\label{sigmad}
  \sigma^2=  \frac{k_BT}{mc^2} \;d^2
  \left(\alpha - \beta\frac{d}{R} \right) \;,
\end{equation}
where
$\alpha=(6/\pi)^{1/3}/(\pi\gamma^{2})$ and $\beta=1/(2\pi\gamma^3)$.
For a close-packed lattice we have
$\alpha=0.31345$, $\beta=0.11254$.
Eq.~(\ref{sigmad}) is the variance of the probability distribution of
the relative displacement $\Delta {\bf u}$ of an atom at a distance $R$
with respect to the atom at the origin,
for one phonon polarization~$\lambda$.
Being the sum of many independent modes,
the probability distribution can be taken as Gaussian.
Upon summing over the three polarizations~$\lambda$,
the distribution is also isotropic. 
We consider the distance vector~${\bf r}$ between the atom
at site ${\bf R}$ and the atom at site ${\bf 0}$. It is given
by ${\bf r}= {\bf R}+\Delta {\bf u}({\bf R)}$.
Its probability distribution is a three-dimensional Gaussian
centered at ${\bf R}$,
\begin{equation}\label{Pr}
P({\bf r},{\bf R})=\frac{1}{(\sqrt{2\pi}\sigma)^3}\exp\left(-\frac{({\bf r}-{\bf R})^2}{2\sigma^2}\right) \;.
\end{equation}
For the PDF we need the corresponding radial distribution,
obtained by multiplying
Eq.~(\ref{Pr}) by $r^2$ and integrating over the angles of ${\bf r}$.
We use spherical coordinates with the $z$-axis along ${\bf R}$.
After some straightforward algebra, we find
the radial probability distribution function
\[
P(r,R) = \frac{1}{\sqrt{2\pi}\;\sigma}\frac{r}{R}\times
\]\vspace{-.5em}
\begin{equation}\label{PrR}
  \left(
  \exp\left({-\frac{(r-R)^2}{2\sigma^2}}\right)
- \exp\left({-\frac{(r+R)^2}{2\sigma^2}}\right) \right) \;,
\end{equation}
where $\sigma$ is a function of $R$, given by Eq.~(\ref{sigmad}).
The PDF of the harmonic crystal at finite temperature is obtained
by replacing the delta-functions
in the zero-temperature PDF, Eq.(\ref{gcrystal0}),
by the peak function~(\ref{PrR}), which yields
\begin{equation}\label{gcrystal}
4\pi r^2\rho\; g(r)  =\sum_{s=1}^\infty n_s P(r,R_s) \;.
\end{equation}
The temperature dependence comes from the peak width $\sigma(R,T)$
given by Eq.~(\ref{sigmad}).
In the following we shall refer to the reduced temperature $T/T_0$,
where $T_0\equiv mc^2/k_B$. Note that in the harmonic approximation
the speed of sound~$c$ is temperature independent and so $T_0$ is a
constant.

\begin{figure}
\begin{center}
\includegraphics[width=1.\columnwidth]{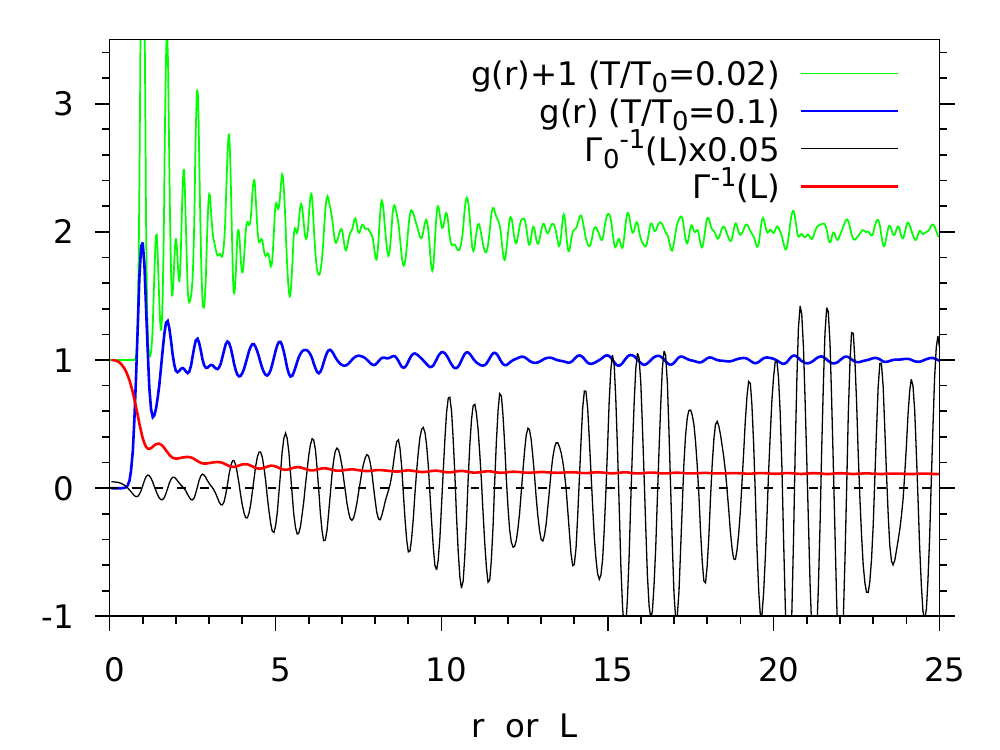}
\caption{PDF $g(r)$ and fluctuation function $\Gamma^{-1}(L)$
  at finite temperature for the fcc crystal in the harmonic approximation.
  The top-most (green) curve is the PDF for
  $T/T_0=0.02$, shifted up by 1 for clarity.
  All other curves are obtained with $T/T_0=0.1$.
  $\Gamma^{-1}$ (red) and  $\Gamma_0^{-1}$ (black)
  correspond to running and finite volume KBI, respectively.
  Note that $\Gamma_0^{-1}$ is multiplied by a factor of 0.05.
}\label{fig2}
\end{center}
\end{figure}
In Fig.~\ref{fig2} the PDF $g(r)$ of Eq.~(\ref{gcrystal})
is plotted for $T/T_0=0.02$ (green line) and $T/T_0=0.1$ (blue line)
for the fcc crystal.
As expected, the PDF becomes smoother with increasing temperature,
but its oscillations do not decay exponentially.
From the PDF for $T/T_0=0.1$, the fluctuation function
is computed with either running or finite volume KBI.
The running KBI $\Gamma_0^{-1}(L)$ (black line) strongly diverges.
Comparison with the $T=0$ result in
Fig.~\ref{fig1} shows that the oscillations are reduced at finite temperature
but they still increase linearly with~$L$.
As a consequence, the running KBI cannot be used for crystals,
even at finite temperature.
In contrast, the result obtained with the finite-volume KBI
($\Gamma^{-1}(L)$, red line) converges smoothly to a finite, positive
limit $\Gamma^{-1}(\infty)$.

\begin{figure}
 \begin{center}
\includegraphics[width=1.\columnwidth]{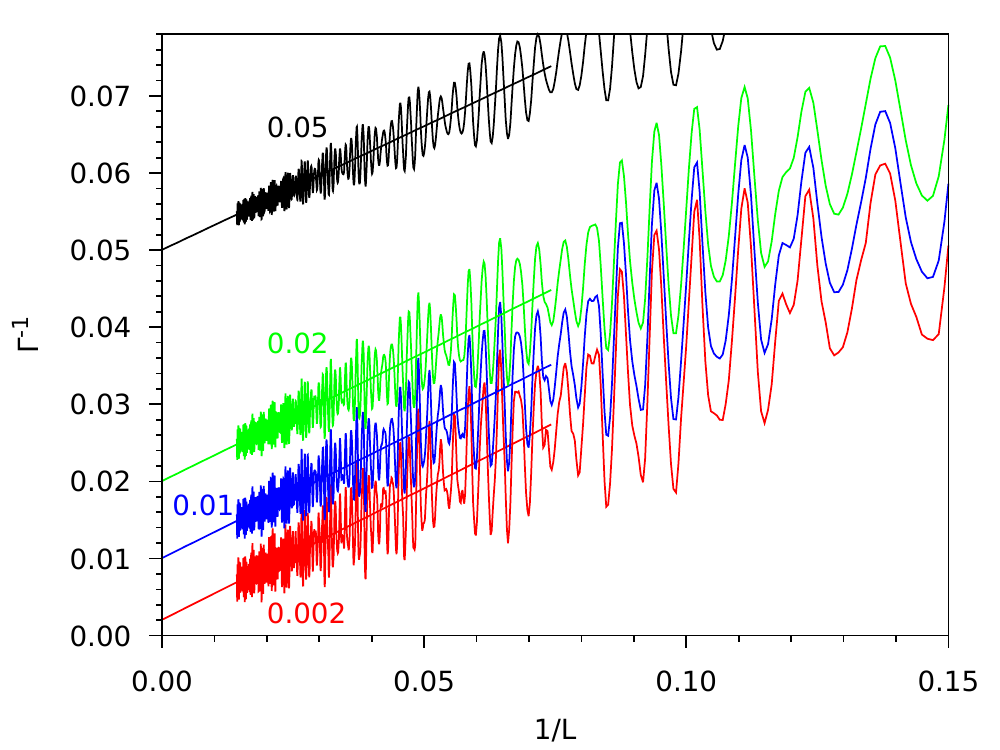}
\caption{$L\rightarrow\infty$ convergence of the
  fluctuation function~$\Gamma^{-1}(L)$ of the harmonic
  fcc crystal obtained with finite-volume KBI.
  The reduced temperature $T/T_0$ is indicated
  on each curve. The straight lines are linear fits in the range
  $1/L<0.075$.}\label{fig3}
\end{center}
\end{figure}
\begin{figure}
 \begin{center}
\includegraphics[width=1.\columnwidth]{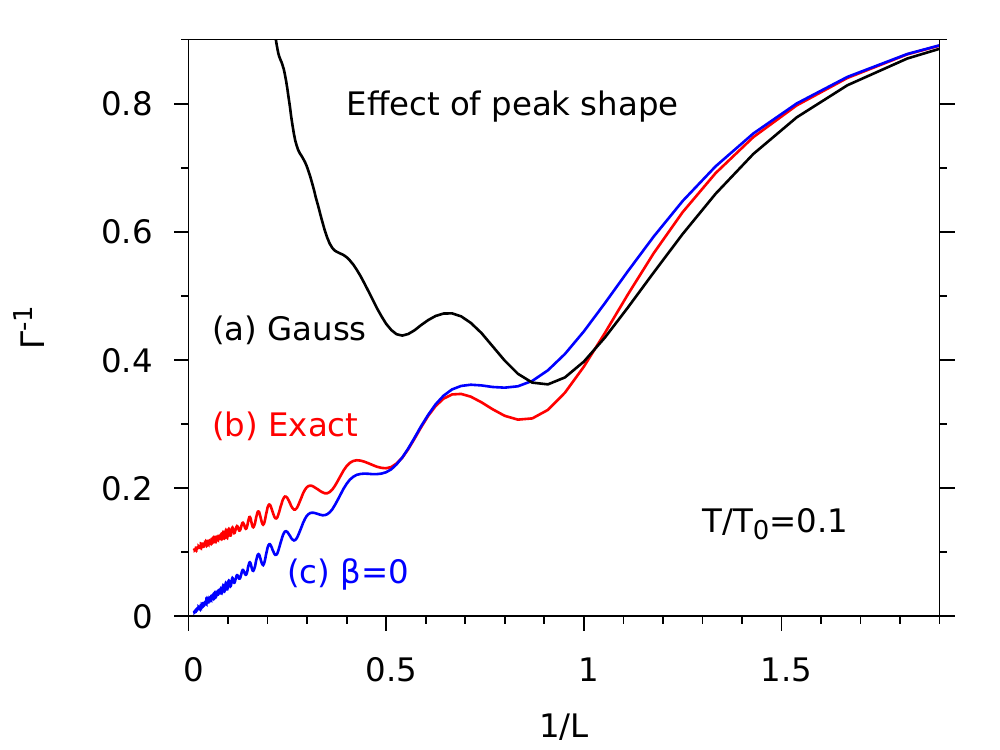}
\caption{
  Effect of peak shape on fluctuation function~$\Gamma^{-1}(L)$ of an fcc crystal at $T/T_0=0.1$, obtained with finite-volume KBI.
  (a) Simple Gaussian shape.
  (b) Exact shape and width (Eqs~\protect\ref{sigmad},\ref{PrR}).
  (c) Exact shape, but constant width ($\beta=0$ in Eq.~\protect\ref{sigmad}).
}\label{fig4}
\end{center}
\end{figure}

In Fig.~\ref{fig3} the convergence of $\Gamma^{-1}(L)$
is studied as a function of $1/L$ for reduced temperatures $T/T_0$ 
in the range from 0.002 to 0.05. Assuming $T_0=5000$~K,
which is a typical value for solid argon, this corresponds to
$T$=10--250~K.
It can be seen that at all temperatures, the oscillations are large,
but the curves clearly converge to $\Gamma^{-1}(\infty)=T/T_0$.
This is the value predicted by the compressibility equation
as will be shown below.
The numerical values of $\Gamma^{-1}(\infty)$,
obtained from a linear regression of the curves in the range $1/L<0.075$
(straight lines in Fig.~\ref{fig3}) are
0.05003, 0.02005, 0.01005, 0.00206.
With respect to the theoretical value $T/T_0$ the absolute error is
about $5\times 10^{-5}$ in all cases and the relative error varies between 0.06\% and 3\%.
It is interesting to note that the correct temperature dependence of the
KBI depends crucially on using the exact form of the peak
shape, Eq.~(\ref{PrR}) and peak width, Eq.~(\ref{sigmad}).
In Fig.~\ref{fig4} the fluctuation function~$\Gamma^{-1}(L)$ is computed
for slightly modified peak shapes.
First, if the peak function~(\ref{PrR}) is
replaced by a simple, normalized Gaussian of the same width, then
$\Gamma^{-1}$ diverges (Fig.~\ref{fig4}~(a), black line).
Second, if the exact peak shape Eq.~(\ref{PrR}) is used but with a
fixed peak width, i.e. $\beta=0$ in Eq.~(\ref{sigmad}),
then $\Gamma^{-1}(L)$ converges to zero instead of $T/T_0$
for $L\rightarrow\infty$ (Fig.~\ref{fig4}~(c), blue line).
At finite temperature, $\Gamma^{-1}(\infty)=0$ implies vanishing
compressibility, which is unphysical.
Comparison between Fig.~\ref{fig4} (b) and (c)
also shows that the temperature dependence of the KBI
is entirely due to the small variation of the PDF peak width with
distance, i.e. the $1/R$ term in Eq.~(\ref{sigmad}).

From the results in Fig.~\ref{fig3} we conclude that the
finite-volume KBI of the harmonic crystal converges to
$\Gamma^{-1}(\infty)= T/T_0$ for any temperature,
where $T_0=mc^2/k_B$ by definition. The speed of sound~$c$ is related to
the isentropic compressibility $\kappa_S$, by the Newton-Laplace equation,
$\kappa_S=(\rho m c^2)^{-1}$, which holds in any phase of matter.
Thus we have $\Gamma^{-1}(\infty)= \kappa_S \rho k_B T$.
In condensed matter, $\kappa_S\approx\kappa_T$ where
$\kappa_T$ is the isothermal compressibility.
Equality holds exactly when the thermal expansion is zero,
which is the case for harmonic interaction considered here~\cite{ashcroft}.
With Eq.~(\ref{gamma}),
it follows that $1+\rho G^\infty\equiv\Gamma^{-1}(\infty)=\kappa_T \rho k_B T$.
This is the compressibility equation~\cite{mcquarrie,bennaim,landau}.
We have thus proved that this fundamental relation of the statistical mechanics
of fluids, also holds in crystalline solids.
The only difference between fluids and solids is the way how $G^\infty$ can be computed.
In fluids, the finite-volume KBI $G^V$ in Eq.~(\ref{GV}) converges absolutely and so the infinite
volume limit $G^\infty$ can be obtained with the usual expression, Eq.~(\ref{ginf}).
In crystals, where the correlation length is infinite, $G^V$ does not converge absolutely
and so the order of integration in Eq.~(\ref{GV}) cannot be changed at will.
As a consequence, the standard expression of $G^\infty$, Eq.~(\ref{ginf}) is ill-defined and
the running KBI~(\ref{G0L}) cannot be used.
Instead, $G^\infty$ must be calculated with the finite-volume KBI, $G^V$
(Eq.~\ref{GV}) or $G(L)$~(Eq.~\ref{GL}).
Our findings show that KBI theory can be applied to crystals,
and that the compressibility can be obtained form the PDF in the same way as in liquids,
provided that  the finite-volume KBI method~\cite{kruger13} is employed.
Here we have used the harmonic approximation and have disregarded
quantum zero-point motion, in order to obtain analytical results.
Anharmonic effects play an important role in the thermodynamic of solids,
and must be taken into account for comparison with experiment.
Anharmonic and quantum effects will modify the PDF, which may be computed using molecular
simulations~\cite{miyaji21}.
However, this does not change the way how the KBI is obtained from the PDF.

In summary, we have generalized KBI theory, which is widely used in
fluids, to crystalline solids. The divergence of standard KBI has been solved
by using the finite volume KBI theory. For a harmonic crystal, we have
derived an analytic expression for the PDF peak shape and have proved that the
compressibility equation holds exactly. The present findings show that
KBI is fully valid in solids, and thus opens new avenues for the thermodynamic
modeling of structurally complex matter and for solid-liquid phase transitions.

\begin{acknowledgments}
I thank Masafumi Miyaji and Jean-Marc Simon for stimulating discussions.
This work was supported by JSPS KAKENHI Grant Number 19K05383.
\end{acknowledgments}


\begin{thebibliography}{99}
\bibitem{mcquarrie}
D. A. McQuarrie, Statistical Mechanics, Harper\&Row, New York, 1973.
\bibitem{bennaim} 
A. Ben-Naim, Molecular Theory of Solutions, Oxford Univ. Press ({2006}).
\bibitem{guevara20}
G. Guevara-Carrion, R. Fingerhut, and J. Vrabec,
Fick Diffusion Coefficient Matrix of a Quaternary Liquid Mixture by
Molecular Dynamics,
J. Phys. Chem. B {\bf 124}, 4527 (2020).
\bibitem{kb52} J. G. Kirkwood, and F. P. Buff,
  The statistical mechanical theory of solutions. I,
  J. Chem. Phys. {\bf 19}, 774 (1951).
\bibitem{landau}
L. D. Landau, E. M. Lifshitz, Statistical Physics, Part 1, 3rd ed.,
Butterworth-Heinemann, Oxford, 1980.
\bibitem{ashcroft}
N. W. Ashcroft and N. D. Mermin, Solid State Physics,
Holt, Rinehart and Winston, New York, 1976,
Chapter 25.
\bibitem{billinge07}
S. J. L. Billinge and I. Levin,
The Problem with Determining Atomic Structure at the Nanoscale,
Science  {\bf 316}, 561-565 (2007).
\bibitem{toby}
  B. H. Toby, T. Enami, Accuracy of Pair Distribution Function Analysis Applied
  to Crystalline and Non-Crystalline Materials, Acta Cryst. A {\bf 48}, 336 (1992).
\bibitem{proffen05}
T. Proffen, K. L. Page. S. E. McLain, B. Clausen, T. W. Darling,
J. A. TenCate, S.-Y. Lee, E. Ustundag,
Atomic pair distributition function analysis of materials
containing crystalline and amorphous phases,
Z. Kristallogr. {\bf 220}, 1002-1008 (2005).
\bibitem{shulgin06} I. L. Shulgin and E. Ruckenstein,
The Kirkwood-Buff Theory of Solutions and the Local Composition of
Liquid Mixtures, J. Phys. Chem. B {\bf 110}, 12707 (2006).
\bibitem{milzetti18}
J. Milzetti, D. Nayar, N. F. A.  van der Vegt,
Convergence of Kirkwood-Buff Integrals of Ideal and Nonideal Aqueous
Solutions Using Molecular Dynamics Simulations,
J. Phys. Chem. B {\bf 122}, 5515 (2018).
\bibitem{kruger13} P. Kr\"uger, S. K. Schnell, D. Bedeaux, S. Kjelstrup,
  T. J. H. Vlugt and J.-M. Simon,
  Kirkwood-Buff integrals for finite volumes
  J. Phys. Chem. Lett. {\bf 4}, 235 (2013).
\bibitem{kruger20}
P. Kr\"uger,
Ensemble averaged Madelung energies of finite volumes and surfaces,
Phys. Rev. B {\bf 101}, 205423 (2020).
\bibitem{schnell11}
  S. K. Schnell, X. Liu, J.-M. Simon, A. Bardow, D. Bedeaux, T. J. H. Vlugt,
  S. Kjelstrup,
  Calculating thermodynamic properties from fluctuations at small scales,
  J. Phys. Chem. B {\bf 115}, 10911 (2011).
\bibitem{dawass18}
N. Dawass, P. Kr\"uger, S. K. Schnell, D. Bedeaux, S. Kjelstrup, J.-M. Simon and T. J. H. Vlugt,
Finite-size effects of Kirkwood-Buff integrals from molecular simulations,
  Mol. Sim. {\bf 44}, 599 (2018).
\bibitem{kruger18}
P. Kr\"uger and T. J. H. Vlugt,
Size and shape dependence of finite-volume Kirkwood-Buff integrals,
Phys. Rev. E {\bf 97}, 051301(R) (2018). 
\bibitem{miyaji21}
M. Miyaji, B. Radola, J.-M. Simon and P. Kr\"uger,
Extension of Kirkwood-Buff theory to solids and its application to the compressibility of fcc argon,
J. Chem. Phys. {\bf 154}, 164506 (2021).
\end{thebibliography}
\end{document}